\documentclass[aps,preprint]{revtex4}%
\usepackage{amsfonts}
\usepackage{amsmath}
\usepackage{amssymb}
\usepackage{graphicx}%
\begin{document}
\preprint{ }
\title{Demonstration of an erbium doped microdisk laser on a silicon chip}
\author{T.~J.~Kippenberg}
\affiliation{and Max Planck Institute of Quantum Optics, Garching, Germany.}
\author{J.~Kalkman and A.~Polman}
\affiliation{Center for Nanophotonics, FOM-Institute AMOLF, Amsterdam, The Netherlands.}
\author{K.~J.~Vahala}
\affiliation{Department of Applied Physics, California Institute of Technology, Pasadena, USA.}

\begin{abstract}
An erbium doped micro-laser is demonstrated utilizing $\mathrm{SiO_{2}}$
microdisk resonators on a silicon chip. Passive microdisk resonators exhibit
whispering gallery type (WGM) modes with intrinsic optical quality factors of
up to $6\times{10^{7}}$ and were doped with trivalent erbium ions (peak
concentration $\mathrm{\sim3.8\times{10^{20}cm^{-3})}}$ using MeV ion
implantation. Coupling to the fundamental WGM of the microdisk resonator was
achieved by using a tapered optical fiber. Upon pumping of the $^{4}%
I_{15/2}\longrightarrow$ $^{4}I_{13/2}$ erbium transition at 1450 nm, a
gradual transition from spontaneous to stimulated emission was observed in the
1550 nm band. Analysis of the pump-output power relation yielded a pump
threshold of 43 $\mathrm{\mu}$W and allowed measuring the spontaneous emission
coupling factor: $\beta\approx1\times10^{-3}$.

\end{abstract}
\maketitle

The increasing demand in computing power and communication bandwidth has
generated an increased interest in the field of silicon photonics which aims
at creating photonic elements utilizing standard, complementary metal oxide
semiconductor (CMOS) processing technology and materials, such as silica and
silicon. As silica and silicon intrinsically lack direct optical transitions,
alternative methods such as erbium doping or creating $\mathrm{Si-SiO_{2}}$
nanostructures \cite{Min1996,Zacharias2002} have been used to achieve optical
emission. Despite these advances, achieving lasing in CMOS compatible
structures has remained challenging and has only been observed recently via
the Raman nonlinearity \cite{Kippenberg2004,Rong2005} and by Er doping of
silicon chip-based silica toroid micro-cavities \cite{Polman2004}.

In this context, erbium is a particularly promising optical dopant as it
provides gain in the 1.55 $\mu$m telecommunication range, and can be
incorporated into a $\mathrm{SiO_{2}}$ by ion implantation \cite{Polman1997},
which is an inherently CMOS compatible process. However, due to the small
emission cross section of Er at 1.55 $\mathrm{\mu m}$ $\mathrm{(\sigma
=4\times10^{-21}cm^{2}}$) in conjunction with the fact that Er concentrations
are limited to $N_{Er}$ $\sim1\times10^{22}$ ions $\mathrm{/cm^{3}}$ due to
clustering, the modal gain is limited to approximately $\sim$7 dB/cm. Lasing
action under these Er gain conditions requires optical resonators with quality
factors $Q>\frac{2\pi}{\lambda}n\frac{1}{\Lambda\sigma_{\varepsilon}N_{Er}%
}\approx10^{5}$ (assuming mode overlap $\Lambda=0.3$ and refractive index
n=1.44). These quality factors are readily available in toroidal
micro-cavities \cite{Armani2003} or silica microspheres \cite{Braginskii1990},
which use a laser assisted reflow process to create ultra-high Q cavities.
Indeed, using these microcavity geometries rare-earth doped microcavity lasers
were first demonstrated \cite{Polman2004,Yang2003,Sandoghdar1996}. However,
the use of a $\mathrm{CO_{2}}$ laser reflow makes control through ion
implantation difficult, since restructuring of the silica takes place.
Increased control of the Er distribution relative to the optical mode is
essential to achieve low lasing threshold or high gain. In addition, increased
control is important in more complex materials systems such as e.g. Er-doped
silica co-doped with Si nanocrystals that act as sensitizers for Er
\cite{Kik2000}.

A more amenable geometry to these studies are planar microdisks
\cite{McCall1992,Kippenberg2003}, which can be fabricated with small
transverse dimensions on a Si chip. Fabrication of these disks does not rely
on a laser reflow process and doping with rare earth ions or Si ions by ion
implantation can be readily performed. Earlier work has already demonstrated
that $Q>10^{6}$ (at 1550 nm) can be achieved in silica microdisks
\cite{Kippenberg2003}, which indicates the possibility to observe Er lasing in
$\mathrm{SiO_{2}}$ microdisks.

In this Letter we demonstrate lasing in an Er-doped microdisk on a silicon
chip. These micro-lasers combine modal engineering of microdisk resonators
with the nanoscale precise control of the Er ion distribution in the disk
through ion implantation, yielding optimum overlap of the Er ions with the
fundamental whispering gallery modes (WGM). By optical pumping of the Er ions
at 1.48 $\mu$m via a tapered optical fiber \cite{Spillane2003}, lasing at 1.55
$\mu$m was observed to occur at a threshold power of less than 43 $\mu$W.
These results demonstrate Er lasing from a microdisk resonator for the first
time, using CMOS compatible fabrication. Fabrication of Er-doped microdisk
lasers proceeded in several steps. The substrate used in the present
experiments was a Si(100) wafer covered with 1 $\mu$m thick thermally-grown
$\mathrm{SiO_{2}}$ film. This oxide thickness represents a compromise between
optical cavity design and ion distribution. The Er ions were incorporated in
the $\mathrm{SiO_{2}}$ by 2 MeV ion implantation at room temperature. The
corresponding implantation range of 560 nm was chosen to obtain a good modal
overlap with the fundamental whispering gallery modes (WGM)(cf. Fig. 1). A
total Er fluence of $4.2\times10^{15}$ ions/$\mathrm{cm^{2}}$ ions was
implanted, yielding a Gaussian depth distribution with a standard deviation
$\sigma\approx72 nm$. The average Er density within the implanted layer is
$N_{Er}=3.8\times10^{20}$ ions/$\mathrm{cm^{3}}$. (integrated over the
full-width at half maximum of the distribution), which corresponds to a modal
gain of 2.66 $\mathrm{cm^{-1}}$ for {$\sigma_{e}=7\times10^{-21}
\mathrm{cm^{2}}$}. Upon implantation the oxidized wafer was annealed in Ar for
1 hour at 800 C, which yielded optimized photoluminescence intensity and
lifetime. The lifetime found ($\sim$14 ms) demonstrated successful passivation
of implantation induced defects. In addition, a reference sample was
fabricated in which the Er implantation at an energy of 4 MeV led to an
implantation depth exceeding the oxide thickness of 1000 nm. While this
implantation depth precludes the observation of lasing, it served as a
reference to assess any deterioration of the Q factor due to ion implantation.

Following Er ion implantation, microcavity fabrication was carried out by
first defining circular silica microdisks using optical lithography and
hydrofluoric etching as detailed in Refs.\cite{Armani2003,Kippenberg2003}. The
resulting microdisk had a diameter of 60 $\mu$m as shown in the scanning
electron micrograph (SEM) in Fig. 1. A key feature of the $\mathrm{SiO_{2}}$
disk, as seen in Fig. 1, is the strong inclination of the cavity sidewalls
that is inherent to the fabrication process, which employs an isotropic HF
etch. The disks were then undercut using a $\mathrm{XeF_{2}}$ gas to
isotroptically etch the silicon and thereby create an air-clad whispering
gallery mode structure. To optically test the microcavities, tapered optical
fibers were used which provide high coupling ideality \cite{Spillane2003}.

First, the reference microdisk resonators were tested in which no lasing of Er
is expected. In these experiments the cavity modes typically appeared as
doublets in the transmission spectrum, which is well known to result from
scattering-induced coupling of the clockwise and counterclockwise cavity modes
\cite{Kippenberg2002,Weiss1995}. An example of such a transmission spectrum is
plotted as an inset in Fig. 2. Upon fitting the data the inferred Q-factor for
each of the two modes was $5\times10^{7}$. This is a very high value for a
planar microdisk, but still one order of magnitude lower than in the case of
toroid microcavities \cite{Armani2003}. Importantly, these Q-factors should
readily allow for observation of lasing of the Er ions. The observed high Q is
attributed to the wedged-shaped edge of the disk microcavity, which is
believed to isolate modes from the disk perimeter and thereby reduce
scattering losses \cite{Kippenberg2003}. This conjecture is further
corroborated using numerical finite-element simulations as shown in Fig. 1,
which demonstrate that an increased sidewall angle leads to an optical mode
that is progressively more removed from the outer, lower cavity boundary
(which can induce scattering losses). The Q value of $5\times10^{7}$ also
provides a lower bound on the effect of ion implantation induced defects.

Next, the 2 MeV Er-implanted microdisks, with an active Er distribution
peaking at a depth of 560 nm were analyzed. First, the Q-factors in the range
from 1410-1480 nm were measured for a microdisk with a diameter of 60 $\mu$m
(and an equivalent free-spectral-range, FSR=9.1 nm) as shown in Fig. 2. To
avoid variations in the overlap factor and population-dependent Q, care was
taken to measure the Q at the same launched power for the same mode family
(fundamental microdisk WGM). Upon approach of the $\mathrm{Er^{3+}}$
absorption band an increase in loss is evidenced. Specifically, while at 1428
nm the quality factor was observed to be $1.5\times10^{6}$ it gradually
reduced to a value as low as $2\times10^{5}$ at 1470 nm. Both the absolute
value of the Q factor as well as its spectral dependence as shown in Fig. 2
are in good agreement with the theoretically predicted $\mathrm{Er^{3+}}$
absorption-limited Q (solid red curve in Fig. 2) assuming $\Lambda$=0.3, a
typical value for the fundamental WGM modes of the microdisk (cf. Fig. 1).
Calculations for $\Lambda$=0.2 and $\Lambda$=0.5 are shown for comparison
(dotted lines). The inset of Fig. 2 shows the Q-dependence of the modal
coupling parameter \cite{Kippenberg2002} ($\Gamma=\tau/\gamma$) defined as the
ratio of is the splitting frequency of the degeneracy-lifted cavity modes
($\gamma^{-1}/2\pi$), and the intrinsic cavity linewidth ($\tau^{-1}/2\pi$),
both derived from the transmission spectra as in the inset of Fig. 2. Data are
taken for several fundamental WGM. The linear dependence that is observed
demonstrates that the quality factors are dominated by absorption losses, in
this case by $\mathrm{Er^{3+}}$ (i.e. since the scattering rate is observed to
be nearly constant for all pump modes the variation of $\Gamma$ must be caused
by a change in absorption).

The excited erbium ions within the microcavity readily couple to the cavity
modes. Due to the Stark broadening of~$^{4}I_{15/2} \longrightarrow$
$^{4}I_{13/2}$ transitions of $\mathrm{Er^{3+}}$ in $\mathrm{SiO_{2}}$, the
ions are coupled to many modes of the cavity. Indeed, as shown in Fig 3, upon
pumping at 1450 nm the spectrum collected through the coupled fiber taper
contains several peaks throughout the erbium emission band (separated by the
FSR of 9.1 nm) clearly demonstrating that the erbium ions are coupled to the
(fundamental) cavity modes. Weaker, subsidiary peaks observed in Fig. 3 are
attributed to Er ions coupling to fundamental WGM's of opposite polarization.
Note that relative strength of the observed $\mathrm{Er^{3+}}$ peak emission
collected in the tapered fiber depends both on the $\mathrm{Er^{3+}}$ emission
spectrum, and the wavelength-dependent ratio of internal to external quality
factor, and therefore does not correspond exactly to the $\mathrm{Er^{3+}}$
emission cross section spectrum. Upon increasing the launched pump power, the
erbium related luminescence in all modes increases linearly (data not shown),
demonstrating that the emission observed in Fig. 3(a) is due to spontaneous emission.

Upon further increase in launched pump power super-linear behavior in the
spontaneous emission is observed, as plotted in Fig. 4, followed by a linear
pump-output relationship of one of the modes at high power (cf. Fig. 3(b)),
which we attribute to lasing. This change is accompanied by an increase in
differential slope efficiency for the lasing mode. The remaining (non-lasing)
modes did not show this transition. This behavior is well known in microcavity
lasers, which due to a large spontaneous emission coupling factor exhibit a
gradual transition from spontaneous emission to stimulated emission
\cite{Yokoyama1989}. From linear interpolation the lasing threshold was
estimated to be 43 $\mu$W. This threshold value is consistent with the
predicted value\cite{Min2004} using V$_{m}=2\pi RA_{m}\approx600\left(
\frac{\lambda}{n}\right)  ^{3},$ $N_{T}=3.8\times10^{20}$cm$^{-3},$
$\lambda_{p}=1480$ nm, $\lambda_{s}=1555$ nm, $\Lambda=0.3$ and assuming that
the intrinsic Q-factor of the lasing mode is $Q_{s}=6\times10^{5}$. The
emission was observed to be single-mode for pump powers up to 400 $\mu$W, and
the highest output power observed was 10 $\mu$W. As we have demonstrated for
toroidal cavities \cite{Min2004} higher output powers can be achieved for
overcoupled conditions, but at the expense of a higher pump threshold.

To determine the spontaneous emission coupling factor ($\beta$), defined as
the fraction of spontaneous emission coupled into the lasing mode with respect
to the spontaneous emission into all modes, the input-output power
relationship was modeled using a rate equation model \cite{Yokoyama1989},
which yields
\begin{equation}
s=\frac{-1}{2\beta}+\frac{p}{2\Omega}+\frac{\sqrt{(\beta p-\Omega)^{2}%
+4p\beta^{2}\Omega}}{2\Omega\beta}%
\end{equation}
Here $s$ is the cavity photon number, $p$ the pump rate (of Er ions) $\Omega$
and the cavity loss rate. The solid line in Fig. 4 is a three parameter fit
using in Eqn. (1), fitting s to the measured output power, $\beta$, and
$\Omega$ . The pump coupling efficiency from fiber into the cavity is assumed
unity. The fit exhibits satisfactory agreement with the data. It is noted
however, that this model can deviate from the observed dependency pump-output
relations close to the threshold, for several reasons. First, the evanescent
waveguide coupling renders the coupled power sensitive to the intrinsic cavity
Q, which due to the presence of Er is varying. Such a pump power-dependent Q
(due to a pump-induced reduction of Er absorption or nonlinear effects), has
already been observed in Raman micro-cavity lasers
\cite{Kippenberg2004,Min2003} . Specifically the coupled power ($\Delta P$) is
given by: $\Delta P=(1-(\frac{Q_{i}-Q_{ex}}{Q_{i}+Q_{ex}})^{2})P$ where $P $
denotes the launched power and $Q_{i}=\left(  1/Q_{0}+1/Q_{s}^{Er}\right)
^{-1}$ is the total quality factor at the pump wavelength, which contains a
contribution from Erbium absorption ( $Q_{s}^{Er}$ ) and other cavity loss
mechanisms ($Q_{0}$ ). The former is given by $1/Q_{p}^{Er}=$ $\frac{\lambda
}{2\pi n}\left(  -\frac{N_{2}}{N_{T}}(\alpha_{p}+g_{p})+\alpha_{p}\right)  $
where $(\alpha_{p}\equiv\Lambda N_{T}\sigma_{p}^{e},g_{p}\equiv\Lambda
N_{T}\sigma_{p}^{a})$ are the Gilles factors describing gain and loss at the
pump wavelength[18] and the normalized upper state population is given (below
threshold) by $\frac{N_{2}}{N_{T}}=$ $\frac{\alpha_{p}\phi_{p}}{\phi
_{p}(\alpha_{p}+g_{p})+N_{T}\cdot1/\tau_{Er}}$(where $\phi_{p}$ is the pump
photon flux and $\tau_{Er}$ the erbium upper state lifetime).For low pump
powers $1/Q_{p}^{Er}=$ $\frac{\lambda}{2\pi n}\alpha_{p}$, whereas for high
pump powers the Q increases to $1/Q_{p}^{Er}=$ $\frac{\lambda}{2\pi n}\left(
-\left(  \frac{N_{2}}{N_{T}}\right)  _{thresh}(\alpha_{p}+g_{p})+\alpha
_{p}\right)  $(which can be close to transparency[18]). Since the Er
absorption is observed to influence the total cavity Q (cf. Fig, 2), it is
clear the pumping dependent Q will lead to loading effects. As the observed Q
factors in the Er absorption band are limited by Er absorption for low pump
powers, this effect will thus lead to an increased intrinsic Q. Effects of
varying Q are most prominent around the threshold, since clamping of the
population occurs above threshold, leading to constant Q as is the case for
low pump powers. Thus the low and high power laser dynamics are well captured
by the above model. A second effect that is not taken into account\ by the
above model is a pump power-dependent Er excited state lifetime, due to up
cooperative upconversion between excited Er ions \cite{Polman1997}. To obtain
an improved fit in the region of low pump and high pump power which determines
the value of $\beta$ (and in which the aforementioned effects are negligible),
the pump-output relationship was fitted on a double logarithmic scale as shown
in the inset of Fig. 4. From the model we derive the spontaneous emission
coupling factor, which describes the fraction of spontaneous emission coupled
into cavity modes: $\beta=1\times10^{-3}$.

We ascribe this relatively low value of $\beta$, to a number of effects. First
of all, $\mathrm{Er^{3+}}$ ions in glass exhibit large homogeneous broadening.
As a consequence the erbium population can decay via a large number of cavity
modes. From the spontaneous emission spectrum in Fig. 2(a), we can estimate
that the total number of cavity modes (N) to which the Er ions couple is
$\sim$20 ( $\sim$10 modes at each TE and TM polarization ). Consequently, even
in the case of spontaneous emission taking place only in cavity modes the
$\beta$ which can be expected is $<5\%$ (=1/N). The yet lower experimentally
observed value is ascribed to that fact that spontaneous emission also occurs
into free-space modes (non-cavity modes).

In summary, we have realized an erbium-implanted microlaser using a silica
microdisk on a silicon wafer. A pump threshold as low as 43 $\mu$W is
observed, and the spontaneous emission coupling factor is determined to be
$\sim1\times10^{-3}$ for the lasing mode, in fair agreement with theory. The
disk geometry presents several advantages over previous toroidal geometries as
it enables the direct use of ion implantation, a planar technology, for doping
optical microcavities. The CMOS compatibility of all fabrication step,
including ion implantation, may enable the use of these microdisk lasers in
photonic and opto-electronic components on a Si chip. Furthermore, these
cavities are ideal microlaboratories to study fundamental effects of a broad
range of ion beam doped optical materials.

\subsection{Acknowledgements}

This work was supported by DARPA and the Caltech Lee Center for Advanced
Networking. TJK acknowledges a postdoctoral fellowship from the Caltech Center
of the Physics of Information. The Dutch part of this work is part of the
research program of FOM which is financially supported by NWO. The authors
thank Ali Dabirian from the MPQ for the finite element numerical modeling. \ \ 

\newpage\

\newpage

\begin{figure}[ptb]
\begin{center}
\includegraphics[]{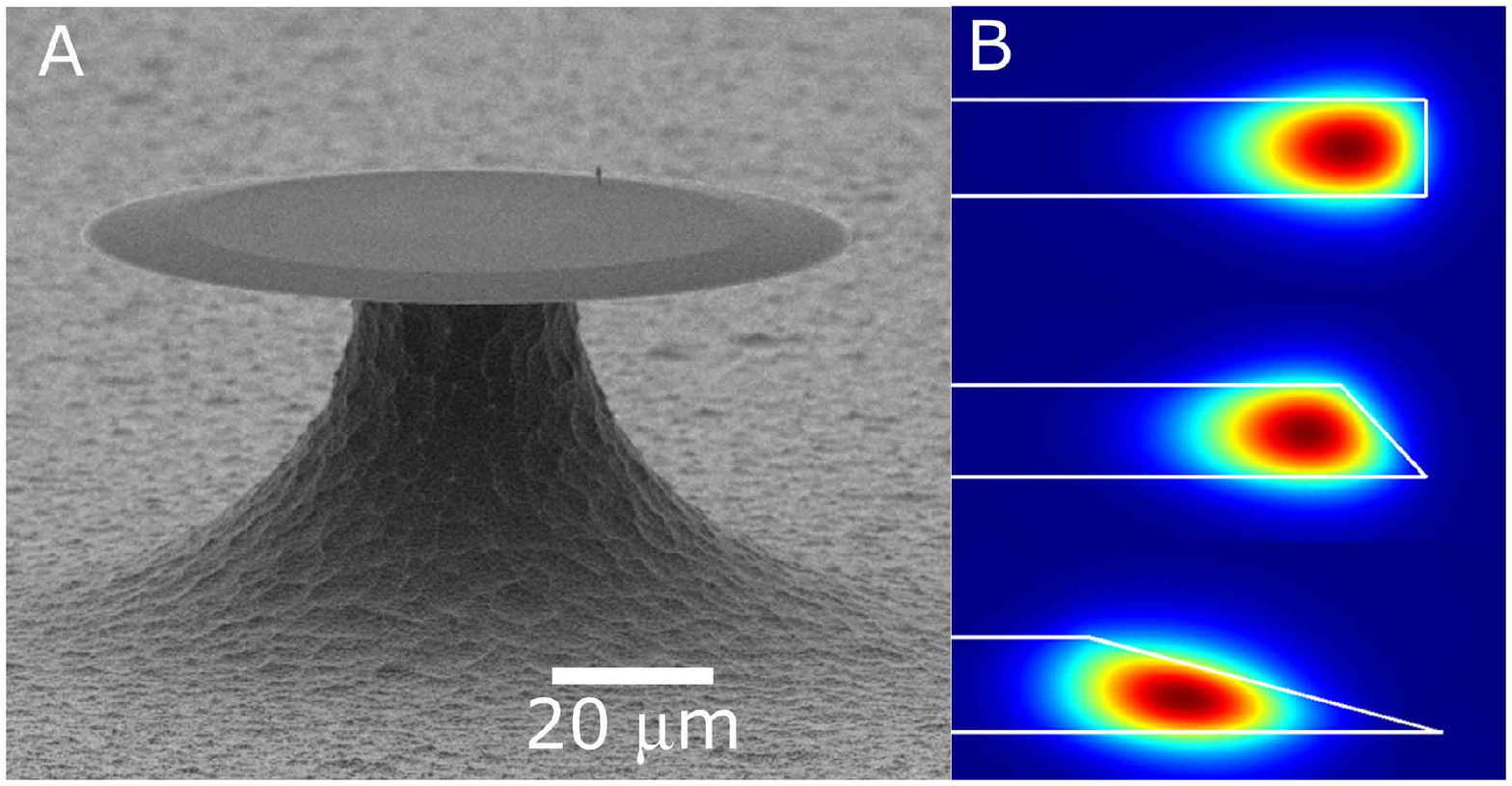}
\end{center}
\caption{ (a) Scanning electron micrograph image of a 60-$\mu$m diameter
silica microdisk cavity on a Si post on a silicon wafer. The wedge shaped
cavity boundary is intentionally induced during fabrication using a
hydrofluoric acid etch. (b) A finite-element simulation of the intensity
profile of the fundamental whispering gallery mode in a microdisk cavity
(oxide thickness 1 $\mu$m, cavity radius 60 $\mu$m, $\lambda=1.54\mu$m) for
three wedge angles: 90, 45, 22 degrees. As can be evidenced in the simulation
a progressive increase in the cavity boundary angle leads to radial shift of
the mode towards the interior of the microdisk, thereby isolating the mode
from the scattering-inducing cavity boundary.}%
\end{figure}

\begin{figure}[ptb]
\begin{center}
\includegraphics[width=3.8in]{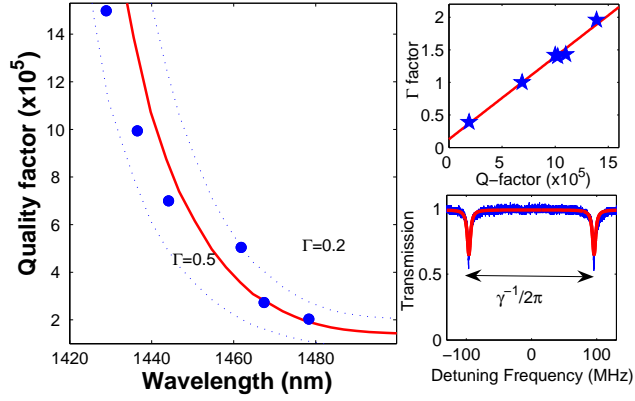}
\end{center}
\caption{Experimentally measured quality factor (Q) for subsequent WGM in the
weak pump regime. All modes belong to the same mode family (fundamental radial
WGM) and are separated by their different angular mode numbers. The solid line
is a calculation taking into account wavelength dependent absorption by Er,
assuming $\Lambda=0.3$. Dotted lines show calculations for $\Lambda=0.2$ and
$\Lambda=0.5$ for comparison. Lower right graph: A transmission spectrum
exhibiting strong mode splitting (from a different sample). The upper right
graph: The modal coupling parameter, i.e. the ratio of the splitting frequency
($\gamma^{-1}/2\pi$) normalized with respect to the intrinsic cavity line
width ($\tau^{-1}/2\pi$), versus Q for the modes measured in the main figure.
The observed linear relationship demonstrates absorption limited Q behavior
(solid line).}%
\end{figure}\newpage

\begin{figure}[ptb]
\begin{center}
\includegraphics[width=3.5in]{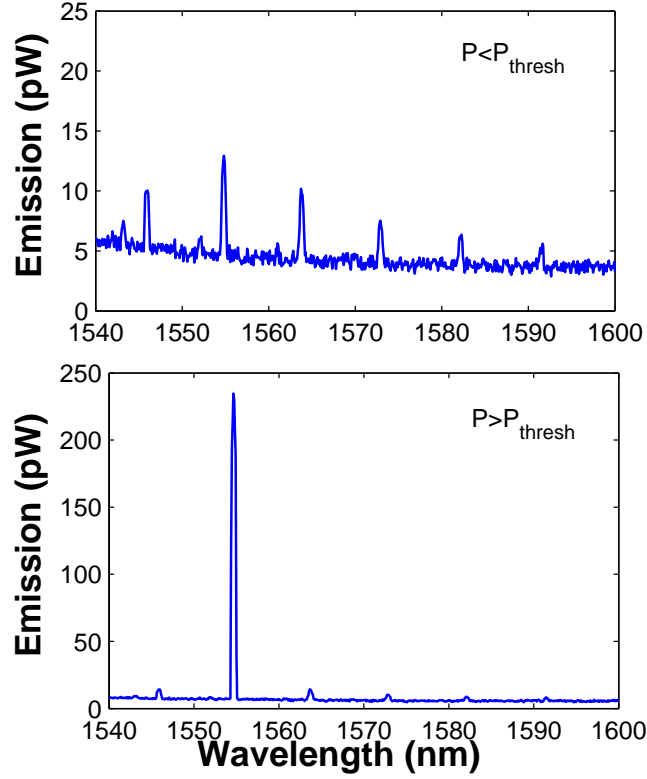}
\end{center}
\caption{Upper graph: The sub-threshold emission collected from the microdisk
through the coupling fiber, when the cavity is resonantly pumped at 1480 nm.
Several fundamental cavity modes are observed (spaced with FSR=9.1 nm). The
weak, subsidiary peaks are attributed to fundamental modes of opposite
polarization. The luminescence maximum at 1555 nm coincides with the peak of
the erbium emission spectrum. Lower graph: The above-threshold spectrum in the
presence of the 1450-nm pump. The non-lasing modes are suppressed by more than
15 dB in comparison with the lasing mode.}%
\end{figure}

\newpage\begin{figure}[ptb]
\begin{center}
\includegraphics[width=3.5in]{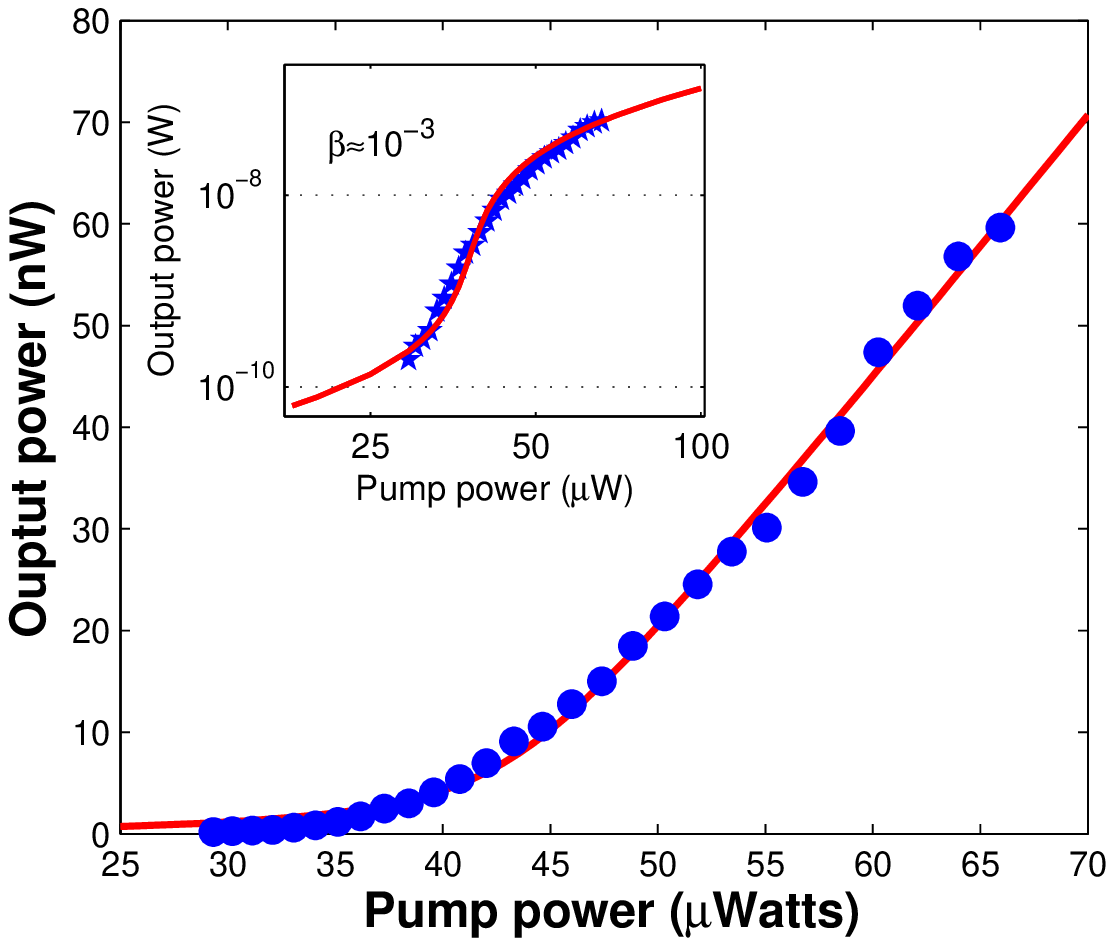}
\end{center}
\caption{Output power versus launched pump power for the Er-doped microdisk in
Fig. 2 (pump wavelength 1480 nm). The transition from spontaneous emission to
stimulated emission is gradual, indicative that a sizeable fraction of the
spontaneous emission is coupled into the micro-cavity. The solid line is a fit
using the model from Ref.\cite{Yokoyama1989} yielding a lasing threshold of 43
$\mu$W. The inset shows the same data on a double-logarithmic scale with a
logarithmic fitting routine which improves fitting for the low power data, and
yields a spontaneous emission coupling factor $\beta\approx1\times10^{-3}$.}%
\end{figure}

\newpage

\newpage

\newpage

\begin{thebibliography}{19}

\bibitem{Armani2003}
D.~K. Armani, T.~J. Kippenberg, S.~M. Spillane, and K.~J. Vahala.
\newblock Ultra-high-q toroid microcavity on a chip.
\newblock {\em Nature}, 421(6926):925--928, 2003.

\bibitem{Braginskii1990}
V.~B. Braginskii, V.~S. Ilchenko, and M.~L. Gorodetskii.
\newblock Optical microresonators with the modes of the whispering gallery
  type.
\newblock {\em Uspekhi Fizicheskikh Nauk}, 160(1):157--159, 1990.

\bibitem{Kik2000}
P.~G. Kik, M.~L. Brongersma, and A.~Polman.
\newblock Strong exciton-erbium coupling in si nanocrystal-doped sio2.
\newblock {\em Applied Physics Letters}, 76(17):2325--2327, 2000.

\bibitem{Kippenberg2003}
T.~J. Kippenberg, S.~M. Spillane, D.~K. Armani, and K.~J. Vahala.
\newblock Fabrication and coupling to planar high-q silica disk microcavities.
\newblock {\em Applied Physics Letters}, 83(4):797--799, 2003.

\bibitem{Kippenberg2004}
T.~J. Kippenberg, S.~M. Spillane, D.~K. Armani, and K.~J. Vahala.
\newblock Ultralow threshold microcavity raman laser on a microelectronic chip.
\newblock {\em Optics Letters}, 2004.

\bibitem{Kippenberg2002}
T.~J. Kippenberg, S.~M. Spillane, and K.~J. Vahala.
\newblock Modal coupling in traveling-wave resonators.
\newblock {\em Optics Letters}, 27(19):1669--1671, 2002.

\bibitem{McCall1992}
S.~L. McCall, A.~F.~J. Levi, R.~E. Slusher, S.~J. Pearton, and R.~A. Logan.
\newblock Whispering-gallery mode microdisk lasers.
\newblock {\em Applied Physics Letters}, 60(3):289--291, 1992.

\bibitem{Min2003}
B.~K. Min, T.~J. Kippenberg, and K.~J. Vahala.
\newblock Compact, fiber-compatible, cascaded raman laser.
\newblock {\em Optics Letters}, 28(17):1507--1509, 2003.

\bibitem{Min2004}
B.~K. Min, T.~J. Kippenberg, L.~Yang, K.~J. Vahala, J.~Kalkman, and A.~Polman.
\newblock Erbium-implanted high-q silica toroidal microcavity laser on a
  silicon chip.
\newblock {\em Physical Review A}, 70(3), 2004.

\bibitem{Min1996}
K.~S. Min, K.~V. Shcheglov, C.~M. Yang, H.~A. Atwater, M.~L. Brongersma, and
  A.~Polman.
\newblock Defect-related versus excitonic visible light emission from ion beam
  synthesized si nanocrystals in sio2.
\newblock {\em Applied Physics Letters}, 69(14):2033--2035, 1996.

\bibitem{Polman1997}
A.~Polman.
\newblock Erbium implanted thin film photonic materials.
\newblock {\em Journal of Applied Physics}, 82(1):1--39, 1997.

\bibitem{Polman2004}
A.~Polman, B.~Min, J.~Kalkman, T.~J. Kippenberg, and K.~J. Vahala.
\newblock Ultralow-threshold erbium-implanted toroidal microlaser on silicon.
\newblock {\em Applied Physics Letters}, 84(7):1037--1039, 2004.

\bibitem{Rong2005}
H.~S. Rong, A.~S. Liu, R.~Jones, O.~Cohen, D.~Hak, R.~Nicolaescu, A.~Fang, and
  M.~Paniccia.
\newblock An all-silicon raman laser.
\newblock {\em Nature}, 433(7023):292--294, 2005.

\bibitem{Sandoghdar1996}
V.~Sandoghdar, F.~Treussart, J.~Hare, V.~LefevreSeguin, J.~M. Raimond, and
  S.~Haroche.
\newblock Very low threshold whispering-gallery-mode microsphere laser.
\newblock {\em Physical Review A}, 54(3):R1777--R1780, 1996.

\bibitem{Spillane2003}
S.~M. Spillane, T.~J. Kippenberg, O.~J. Painter, and K.~J. Vahala.
\newblock Ideality in a fiber-taper-coupled microresonator system for
  application to cavity quantum electrodynamics.
\newblock {\em Physical Review Letters}, 91(4):art. no.--043902, 2003.

\bibitem{Weiss1995}
D.~S. Weiss, V.~Sandoghdar, J.~Hare, V.~Lefevreseguin, J.~M. Raimond, and
  S.~Haroche.
\newblock Splitting of high-q mie modes induced by light backscattering in
  silica microspheres.
\newblock {\em Optics Letters}, 20(18):1835--1837, 1995.

\bibitem{Yang2003}
L.~Yang, D.~K. Armani, and K.~J. Vahala.
\newblock Fiber-coupled erbium microlasers on a chip.
\newblock {\em Applied Physics Letters}, 83(5):825--826, 2003.

\bibitem{Yokoyama1989}
H.~Yokoyama and S.~D. Brorson.
\newblock Rate-equation analysis of microcavity lasers.
\newblock {\em Journal of Applied Physics}, 66(10):4801--4805, 1989.

\bibitem{Zacharias2002}
M.~Zacharias, J.~Heitmann, R.~Scholz, U.~Kahler, M.~Schmidt, and J.~Blasing.
\newblock Size-controlled highly luminescent silicon nanocrystals: A sio/sio2
  superlattice approach.
\newblock {\em Applied Physics Letters}, 80(4):661--663, 2002.

\end{thebibliography}
\end{document}